\documentclass[fleqn,usenatbib]{mnras}
\UseRawInputEncoding

\usepackage{newtxtext,newtxmath}
 
\usepackage{graphicx}
\usepackage[T1]{fontenc}
\usepackage{pdflscape}

\DeclareRobustCommand{\VAN}[3]{#2}
\let\VANthebibliography\thebibliography
\def\thebibliography{\DeclareRobustCommand{\VAN}[3]{##3}\VANthebibliography}

\usepackage{graphicx}	
\usepackage{amsmath}	
\usepackage{amssymb}	

\title[Dusty lensed system at $z\geq$6]{ALMA Lensing Cluster Survey: a strongly lensed multiply imaged dusty system at $z\geq$6}

\author[N. Laporte et al.]{
N. Laporte,$^{1,2}$\thanks{E-mail: nl408@cam.ac.uk}
A. Zitrin,$^{3}$
R. S. Ellis, $^{4}$
S. Fujimoto, $^5$
G. Brammer, $^5$
J. Richard, $^6$
\newauthor
M. Oguri,$^{7,8,9}$
G. B. Caminha, $^{10}$
K. Kohno, $^{11,9}$
Y. Yoshimura, $^{11}$
Y. Ao, $^{14,15}$
F. E. Bauer, $^{19,20}$
\newauthor
K. Caputi, $^{10,5}$
E. Egami, $^{23}$
D. Espada, $^{16,21,22}$
J. Gonz\'alez-L\'opez $^{25,26}$
B. Hatsukade, $^{11}$
\newauthor
K. K. Knudsen, $^{18}$
M. M. Lee, $^{24}$
G. Magdis, $^{5}$
M. Ouchi, $^{7,13}$
F. Valentino, $^{5}$
T. Wang. $^{11,16}$
\vspace{0.1cm} \\
$^{1}$ Kavli Institute for Cosmology, University of Cambridge, Madingley Road, Cambridge CB3 0HA, UK \\
$^2$ Cavendish Laboratory, University of Cambridge, 19 JJ Thomson Avenue, Cambridge CB3 0HE, UK \\
$^3$ Physics Department, Ben-Gurion University of the Negev, P.O. Box 653, Beer-sheva 8410501, Israel\\
$^4$ Department of Physics and Astronomy, University College London, Gower Street, London WC1E 6BT, UK \\
$^5$ Cosmic Dawn Center (DAWN), Copenhagen, Denmark \\
$^6$ Univ Lyon, Univ Lyon1, Ens de Lyon, CNRS, Centre de Recherche Astrophysique de Lyon UMR5574, F-69230, Saint-Genis-Laval,France \\
$^7$ Kavli Institute for the Physics and Mathematics of the Universe (WPI), The University of Tokyo, 5-1-5 Kashiwanoha, Kashiwa-shi, Chiba, 277-8583, Japan \\
$^8$ Department of Physics, University of Tokyo, 7-3-1 Hongo, Bunkyo-ku, Tokyo 113-0033 Japan \\
$^9$ Research Center for the Early Universe, University of Tokyo, Tokyo 113-0033, Japan \\
$^{10}$ Kapteyn Astronomical Institute, University of Groningen,Postbus 800, 9700 AV Groningen, The Netherlands \\
$^{11}$ Institute of Astronomy, Graduate School of Science, The University of Tokyo, 2-21-1 Osawa, Mitaka, Tokyo 181-0015,  Japan \\
$^{13}$ Institute for Cosmic Ray Research, The University of Tokyo, 5-1-5 Kashiwanoha, Kashiwa, Chiba 277€"8582, Japan \\
$^{14}$ Purple Mountain Observatory and Key Laboratory for Radio Astronomy, Chinese Academy of Sciences, Nanjing, China\\
$^{15}$ School of Astronomy and Space Science, University of Science and Technology of China, Hefei, China \\
$^{16}$ National Astronomical Observatory of Japan, 2-21-1 Osawa, Mitaka, Tokyo 181-8588, Japan\\
$^{18}$ Department of Space, Earth and Environment, Chalmers University of Technology, Onsala Space Observatory, SE-43992 Onsala, Sweden \\ 
$^{19}$ Instituto de Astrof\'isica, Facultad de F\'sica, Pontificia Universidad Catolica de Chile Av. Vicu\~na Mackenna 4860, 782-0436 Macul,Santiago, Chile \\
$^{20}$ Millennium Institute of Astrophysics (MAS), Nuncio Monse nor Santero Sanz 100, Providencia, Santiago, Chile\\
$^{21}$ Department of Astronomical Science, SOKENDAI (The Graduate University of AdvancedStudies), 2-21-1 Osawa, Mitaka, Tokyo 181-8588, Japan \\ 
$^{22}$ SKA Organization, Lower Withington, Macclesfield, Cheshire SK11 9DL, UK \\ 
$^{23}$ Steward Observatory, University of Arizona, 933 North Cherry Avenue, Tucson, AZ 85721, USA \\
$^{24}$ Max-Planck-Institut f\"{u}r Extraterrestrische Physik (MPE), Giessenbachstr., D-85748 Garching, Germnay \\
$^{25}$ N\'ucleo de Astronom\'ia de la Facultad de Ingenier\'ia  y Ciencias, Universidad Diego Portales, Av. Ej\'ercito Libertador 441, Santiago, Chile \\
$^{26}$ Las Campanas Observatory, Carnegie Institution of Washington, Casilla 601, La Serena, Chile \\
 }

\date{Accepted XXX. Received YYY; in original form ZZZ}

\pubyear{2015}

\begin{document}
\label{firstpage}
\pagerange{\pageref{firstpage}--\pageref{lastpage}}
\maketitle
\begin{abstract}
We report the discovery of an intrinsically faint, quintuply-imaged, dusty galaxy MACS0600-z6 at a redshift $z=$6.07 viewed through the cluster MACSJ0600.1-2008 ($z$=0.46). A $\simeq4\sigma$ dust detection is seen at 1.2mm as part of the ALMA Lensing Cluster Survey (ALCS), an on-going ALMA Large program, and the redshift is secured via [C II] 158 $\mu$m emission described in a companion paper. In addition, spectroscopic follow-up with GMOS/Gemini-North shows a break in the galaxy's spectrum, consistent with the Lyman break at that redshift.
We use a detailed mass model of the cluster and infer a magnification $\mu\gtrsim$30 for the most magnified image of this galaxy, which provides an unprecedented opportunity to probe the physical properties of a sub-luminous galaxy at the end of cosmic reionisation. Based on the spectral energy distribution, we infer lensing-corrected stellar and dust masses of $\rm{2.9^{+11.5}_{-2.3}\times10^9}$ and $\rm{4.8^{+4.5}_{-3.4}\times10^6}$ $\rm{M_{\odot}}$ respectively, a star formation rate of $\rm{9.7^{+22.0}_{-6.6} M_{\odot} yr^{-1}}$, an intrinsic size of $\rm{0.54^{+0.26}_{-0.14}}$ kpc, and a luminosity-weighted age of 200$\pm$100 Myr. Strikingly, the dust production rate in this relatively young galaxy appears to be larger than that observed for equivalent, lower redshift sources. We discuss if this implies that early supernovae are more efficient dust producers and the consequences for using dust mass as a probe of earlier star formation.

\end{abstract}




\section{Introduction}
Understanding the process of cosmic reionisation, during which intergalactic hydrogen transforms from a neutral to an ionised state in less than a billion years, represents a major challenge in extragalactic astronomy and cosmology. Although star-forming galaxies are considered to be the primary ionising sources, those which can be directly observed during the reionisation era appear to be insufficient in number given their likely ionisation output (e.g. \citealt{Robertson2015}). As a consequence, it is popular to appeal to contributions from a larger number of sub-luminous sources, which can only be partially probed via current deep surveys (e.g., \citealt{Bouwens2015}, \citealt{Livermore2017}, \citealt{Bhatawdekar2019}, \citealt{Kikuchihara2020} ). With current facilities, the only way to characterise the physical properties of this sub-luminous population is to use the magnification afforded by gravitational lensing (for a review see \citealt{Kneib2011}). Although several surveys have harnessed the lensing power of foreground clusters, such as the Cluster Lensing And Supernova survey with Hubble (CLASH ;\citealt{Postman2012}), the \textit{Hubble} Frontier Fields \citep{Lotz2017} and most recently REionization LensIng Cluster Survey (RELICS ;  \citealt{Coe2019}), the effective surface area explored for the most highly-magnified background sources remains small. As a consequence, very few high-redshift sources ($z\geq$6) with magnifications $\mu \geq$10 have so far been found (e.g., \citealt{Coe2013}, \citealt{Salmon2018}). 

Of particular interest in physical studies of galaxies in the reionisation era is their dust content. Although dust can obscure the ionising radiation and lead to underestimates of properties such as star formation rate and ionising capability, dust masses have been proposed as a potential way forward in estimating the earlier star formation histories and hence ages of early galaxies (\citealt{Laporte2017}, \citealt{Katz2019}). It is typically assumed that dust at early times is largely produced in core-collapse supernovae with a production rate similar to that observed in local events ($<$0.2 M$_{\odot}$ per event, e.g. \citealt{Indebetouw2014}). However, little is known about dust production rates at early times and thus correlating dust properties of early galaxies over the full range of stellar masses with stellar ages inferred independently from spectral energy distributions (SEDs) would be highly valuable. To date there are very few convincing dust detections beyond $z\simeq$6 (e.g., \citealt{Capak2015}, \citealt{Watson2015}; \citealt{Bradac2017}; \citealt{Laporte2017}; \citealt{Marrone2018}; \citealt{Hashimoto2019} ; \citealt{Tamura2019}) and none for galaxies with UV magnitude $\rm{M_{uv}>-18}$  

The ALMA Lensing Cluster Survey (ALCS - PI: K. Kohno) is motivated by multiple goals including probing the dust properties of strongly-lensed galaxies near the end of the epoch of reionisation. The programme has observed 33 clusters (5 from the Frontier Fields survey, 12 from CLASH and 16 from RELICS) in ALMA band-6 ($\lambda \sim$1.2mm) to reach a 5$\sigma$ lensing-corrected rms of $\leq$ 0.1 mJy over 16 arcmin$^2$ (or 0.01 mJy over the 0.5 arcmin$^2$ for the highest magnification regions). In this paper, we report the discovery of dust-emission from a multiply-imaged lensed system MACS0600-z6 at $z=$6.07 behind one of the ALCS clusters, MACSJ0600.1-2008 \citep{Ebeling2001}. In Section~\ref{sec.data} we describe the relevant observational data and the photometry of the lensed source. Based on three mass models described in Section~\ref{sec.mass}, we identify five multiple images for this system, infer a total magnification of $\mu\gtrsim$ 30 for the most magnified arc. Section~\ref{sec.spectro} presents spectroscopic follow-up with Gemini-North of several images, and we discuss the physical properties of this system in Section~\ref{sec.properties}. Throughout this paper we assume $\rm{H_0=70 km.s^{-1}.Mpc^{-1}}$, $\Omega_m$=0.3 and $\Omega_{\Lambda}$=0.7. All magnitudes are in AB system \citep{ABsystem}.

\section{Observation and Data Analysis}
\label{sec.data}
The cluster MACSJ0600.1-2008 ($z$=0.46) was observed as part of the ALMA Lensing Cluster Survey (ID: 2018.1.00035.L, P.I. Kohno, K.) in band 6 frequency in January 2019 with a total exposure time of 3.3 hrs. The data was reduced using the ALMA pipeline (v. 5.6.1-8) with a natural weighting. A 2$''$ \textit{uv}-taper was used to maximise the detection of faint sources. The final beam size is 2\farcs\,21 $\times$1\farcs\,99 and the rms on the reduced mosaic is 80 $\mu$Jy/beam. 13 sources with a signal-to-noise (SNR) $\geq$4 were identified including one at RA=06:00:09.143 and DEC=$-$20:08:26.579 with an elongated half light radius of 1\farcs\,3. By extracting the flux on a line-free image at 1.2mm using the CASA \textit{imfit} task, the peak flux is $\rm{S^{peak}_{1.2mm}}$=366$\pm$62 $\rm{\mu Jy.beam^{-1}}$ and the integrated flux is $\rm{S^{int}_{1.2mm}}$=484$\pm$135 $\mu$Jy on an emission line subtracted image.

\begin{figure*}
\includegraphics[width=15cm]{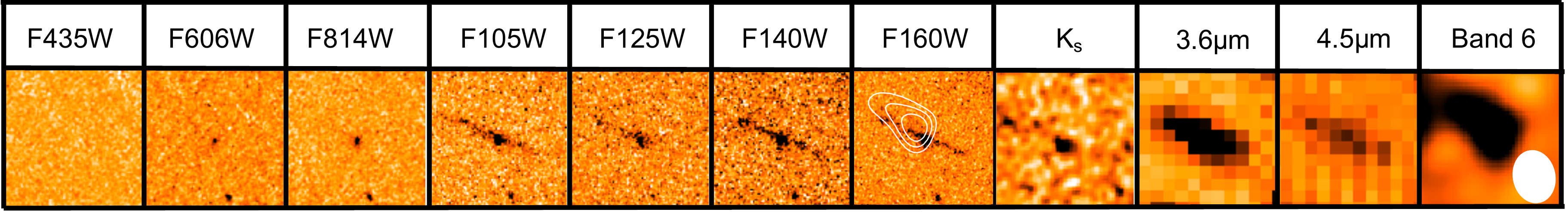}
\caption{\label{stamps} Thumbnail images of MACS0600-arc comprising images from ACS/\textit{HST} (F435W, F606W, F814W), WFC3-IR/HST (F105W, F125W, F140W and F160W), HAWK-I/VLT(K$_s$), IRAC/\textit{Spitzer} (3.6 and 4.5 $\mu$m) and ALMA (Band 6). Each image is 7$''$ $\times$7$''$. Note the overlap between the foreground object (detected at F606W and F814W) and the lensed galaxy. The ALMA contours are drawn on the F160W images in 1$\sigma$ from 3$\sigma$. The ALMA beam size is at the bottom right corner of the last column.}
\end{figure*}

A search for an ultraviolet (UV) rest-frame counterpart for this source was undertaken on deep \textit{HST} images from the RELICS survey with a search radius of 2$''$.  
The data reduction of this dataset is described in \citet{Coe2019}. Photometric catalogues were built using version 2.19.5 of \textit{SExtractor} \citet{SExtractor} in dual image mode on psf matched images. The extraction parameters were defined to maximise the detection of faint objects (DETECT\_THRESH = 2$\sigma$ ; ANALYSIS\_THRESH = 2$\sigma$ ; DETECT\_MINAREA= 5 px) on a detection picture composed by the sum of all WFC3 images. The final catalogue contains 25200 detections. To estimate the depth of each image, we masked the bright sources and measured the rms in hundreds non-overlapping 0.4'' radius aperture. Table~\ref{tab.data} summarises the data properties.
Two overlapping sources were detected 0\farcs \,4 South of the ALMA detection (Figure~\ref{stamps}): a point source (referred to hereafter as `foreground') and a background arc-like source (referred to as MACS0600-z6, whose discovery multiple image is denoted MACS0600-arc, see below for more details). The foreground source was identified in the public RELICS catalogue \citep{Salmon2020} and assigned a photometric redshift of $\rm{z_{phot}}$=0.82$^{+0.28}_{-0.40}$. The multiple image MACS0600-arc is absent from that pipeline-generated catalogue, most likely because of its faintness and elongated shape. 
\begin{table}
    \centering    

   \hspace{-1.4cm}  \begin{tabular}{l|cccc} \hline
Image	&	$\lambda_{c}$ & t$_{exp}$	&	ZP	& 2$\sigma$-depth		\\
    	&	[\AA ] & [ks]	&	[AB]		& [AB]	\\ \hline
F435W & 4317.5 & 1.95   & 25.66 &  27.4  \\
F606W & 5924.7 & 2.18   & 26.50 &  28.3  \\
F814W & 8210.3 & 3.56   & 25.94 &  28.0  \\
F105W & 10530.9 & 1.41   & 26.27 &  27.1  \\
F125W & 12495.7 & 0.7   & 26.23 &  26.5  \\
F140W & 13976.1 & 0.74   & 26.45 &  26.5  \\
F160W & 15433.1 & 1.96   & 25.95 &  27.4  \\ \hline
Ks & 21440.4 & 2.16 & 30.05 & 24.4 \\ \hline
IRAC1 & 35465.6 & 15.2  & 23.9 & 26.0 \\ 
IRAC2 & 45024.3 & 15.2  & 23.9 & 25.6 \\ \hline
    \end{tabular}
    \caption{\label{tab.data} Properties of data used in this study. The 2$\sigma$-depth is measured in a 0.4'' diameter aperture for \textit{HST}, 0.8'' radius aperture for Ks, and 1.2'' radius aperture for \textit{Spitzer}.  }
\end{table}

Noting the overlapping foreground object, in order to extract the photometry of MACS0600-arc we model each source with a Sersic profile using  \textsc{GALFIT} \citep{GALFIT} on the WFC3-NIR and IRAC images, which accounts for PSF difference between HST, VLT and Spitzer. Error bars are measured on the residual image in a 0\farcs\,4 radius circle. We tested our method on the ACS data where only the foreground source is detected and compared the \textsc{GALFIT} photometry with that derived using the standard MAG\_AUTO photometry in single image mode with SExtractor \citep{SExtractor}. In F606W, our extracted photometry ($\rm{m_{F606W}^{GALFIT}=26.55\pm0.35}$) is in excellent agreement with the \textit{SExtractor} photometry ($\rm{m_{F606W}^{SExtractor}=26.72\pm0.08}$). A similar conclusion is seen for the F814W image. 

To increase the wavelength coverage, we searched for deep public data at longer wavelength. Using the ESO archive, we found a deep Ks exposure carried out with HAWKI/VLT (ID: 0100.A-785, P.I. : A. Edge) reaching a 2$\sigma$ depth in 1.2'' radius aperture of 24.4. We also searched in the \textit{Spitzer} archive for deep IRAC 3.6 and 4.5 $\mu$m exposures. This cluster was part of two observing programs (ID: 90218, P.I. : E. Egami and ID: 12005, P.I. : M. Bradac), we combined all exposures and measured the depth in a 1.2'' radius aperture distributed over the field of view.  We present the final photometry in Table~\ref{photometry}.

\begin{figure}
\hspace{-1cm}
\includegraphics[width=11cm]{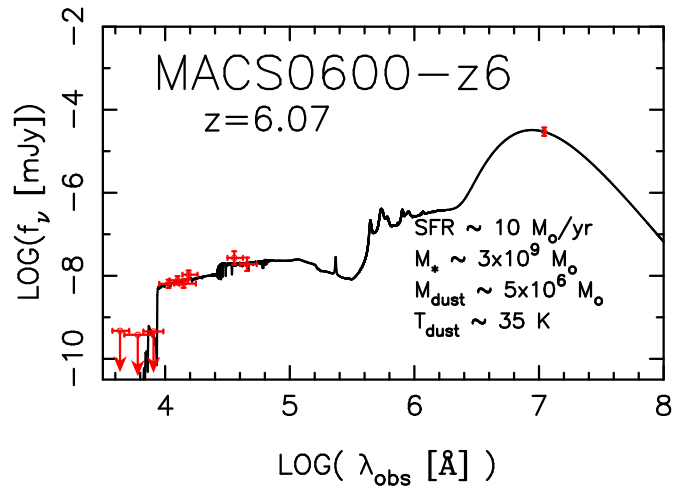}
\caption{\label{SED} Spectral Energy Distribution (SED) of MACS0600-z6 showing the best-fit from MAGPHYS and the best physical properties corrected for magnification. }
\end{figure}

We determine the photometric redshift of the foreground source from a template based SED-fitting method with \textsc{Hyperz} \citep{Hyperz}. We used the standard list of templates including evolutionary SEDs with Chabrier IMF \citep{Chabrier2003} and solar metallicity from \citet{Bruzual2003}, empirical SEDs built by \citet{Coleman1980} and two starburst galaxies from \citet{Kinney1996}. We then searched for a solution between $0<z<8$, with an extinction range $\rm{0.0 < A_v < 3.0}$ mag assuming a Calzetti reddening law \citep{Calzetti2000}. The colour of the foreground source is similar to the colour of cluster members, we therefore allow a redshift range of $z \in$ [0.0:3.0].  The best SED-fit for the foreground source is obtained at $\rm{z_{phot}=0.57^{+0.14}_{-0.17}}$ and modest reddening ($\rm{A_v}$=1.5mag).  Allowing a larger redshift range, the MACS0600-arc data provides a solution at $z_{photo}$=$6.68^{+0.24}_{-0.67}$ with an extinction $\rm{A_v}$=0.5mag, broadly consistent with the spectroscopic value (Figure~\ref{SED}).

\section{Lens models and multiple images}
\label{sec.mass}

The elongated shape of the prominent $z\sim6$ arc supports the suggestion of a strongly lensed source. To determine the lensing configuration and hence the intrinsic properties of this source and the associated uncertainties, we have utilised several different lens models from within the ALCS team. These models exploit spectroscopic observations of MACS0600 using the MUSE integral field unit spectrograph (ID 0100.A-0792, PI: Edge) and the photometric redshift catalogue from RELICS. More details on the lens modelling of MACS0600 will be given in a separate paper (\emph{in preparation}), and here we provide only a brief summary. For convenience in this section we drop the prefix MACS0600 in describing the multiple images.

\begin{figure*}
\includegraphics[width=18cm]{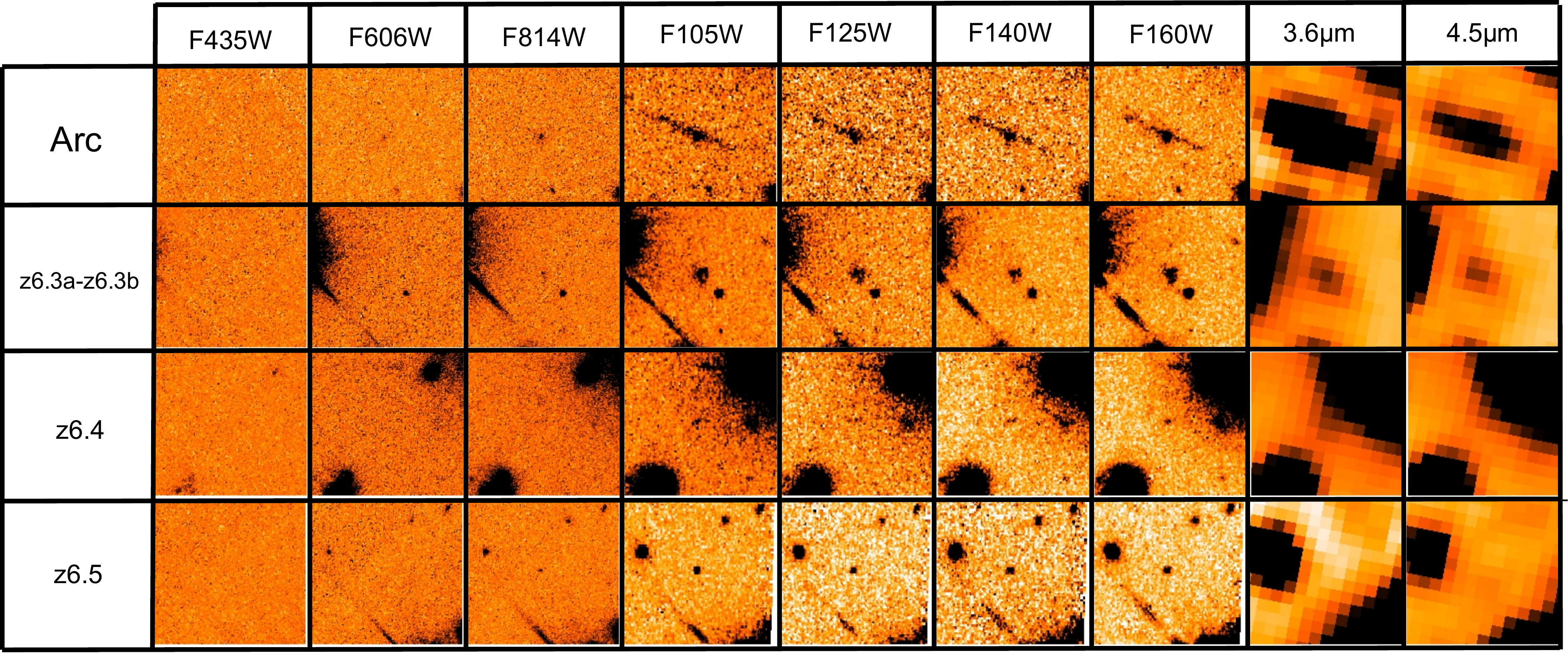}
\caption{\label{stamps-images} Thumbnail images of all images of the quintuply lensed galaxy presented in this paper. Each stamp is 7\farcs5 $\times$ 7\farcs5 where the image is located at the center, excepted for MACS0600-z6.3a (left) and MACS0600-z6.3b (right) where the two images are very closed. For the fifth image, we only show the candidate detected on all WFC3/HST images (see text for details).}
\end{figure*}

\begin{figure*}
\includegraphics[width=18cm]{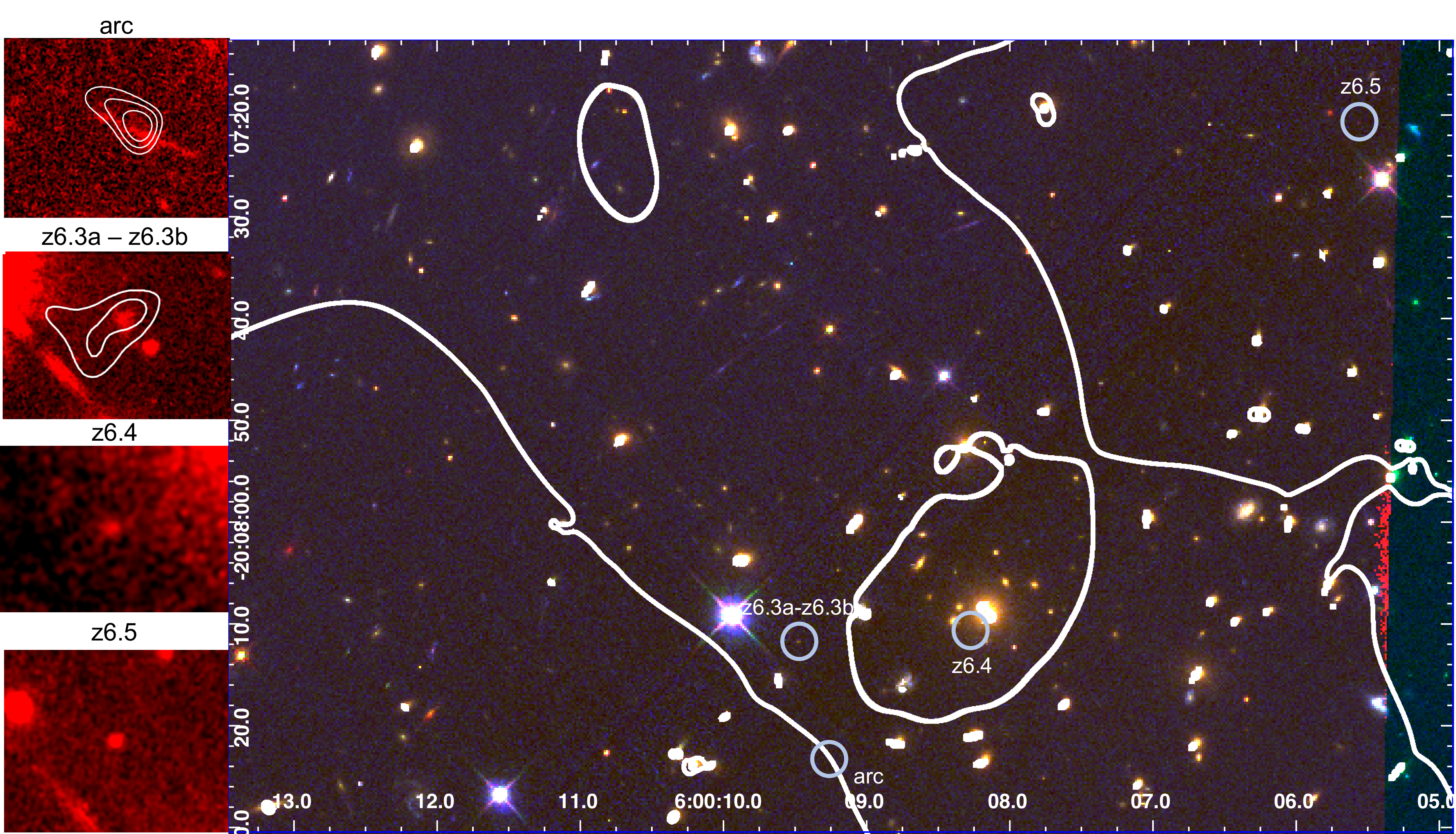}
\caption{\label{images} (\textit{Right}) Position of the five images of MACS0600-z6 superimposed on a colour HST image (F435W - blue ; F814W - green ; F160W - red). The white contours from the LTM model are overplotted. (\textit{Left}) Labelled F160W stamps of four multiple images. MACS0600-z6.1 and MACS0600-z6.2 (top left) are a merging pair crossing the critical line, and seen on HST images as an elongated arc and referred in the text as MACS0600-arc. MACS0600-z6.3 comprises a bright object (z6.3a) and a fainter, fuzzy part (z6.3b), which is the quintuply imaged part as indicated by the lens models, and on which the line emission seems to be concentrated. ALMA contours are overplotted on the arc stamp from $3\sigma$ and from 2$\sigma$ for z6.3b.}
\end{figure*}

The three models used here are produced using \textsc{GLAFIC} \citep{GLAFIC}, \textsc{Lenstool} \citep{Lenstool}, and Light-Traces-Mass (LTM; \citealt{Zitrin2015}). The first two models can be regarded as parametric, in the sense that both the galaxies and their dark matter halos are described using independent analytic forms (e.g., combinations of pseudo isothermal elliptical mass distributions) and adopt empirical scaling relations for cluster members. The LTM method is similar, but here the dark matter is assumed to follow the light distribution and thus is modeled as a smoothed version of the luminosity-weighted galaxy distribution. The parametric modeling of MACS0600 typically included three to four dark matter halos and, whereas for the LTM models, the masses of several bright cluster members were modeled independently. All models are minimized using available multiple image constraints to find the best-fit solution and its associated errors. Four modellers (MO, JR, GC, AZ) voted on the choice of multiply-imaged systems which, besides the $z\sim6$ system, included nine other systems considered as secure. Five of these have a spectroscopic redshift from MUSE (spanning the redshift range $z\sim$1.5 to $z\sim$5.5). All three models indicate that the source MACS0600-z6 is quintuply imaged with the discovery arc (images z6.1 and z6.2) being a pair of multiple images that merge across the critical curve; an additional counter image inside the tangential critical curves (z6.3); a radial image (z6.4); and a fifth image on the other side of the cluster to the west (z6.5). The stamps of the 5 images are shown in Fig. \ref{stamps-images}, and the configuration derived from the LTM model for example, including all five multiple images, is shown on a \textit{HST} image of the cluster in Fig. \ref{images}. To confirm the reliability of this system, we first check that the colors are similar for  the 4 images within the error bars. We only include one of the fifth image candidates in this analysis (z6.5c) because HST coverage is missing for the other two. Colors measured on WFC3 images are similar with averaged values of F125W-F160W = 0.32$\pm$0.15. Another relevant feature that should be conserved by lensing is the surface brightness. We estimated the surface brightness of the 4 images using the extracted magnitude from GALFIT for the arc and z6.3a,b and from our SExtractor catalogues for the remaining objects using the MAG\_AUTO and the isophotal area. All images have consistent surface brightness in the WFC3 images (e.g. the averaged value on F160W is 21.9$\pm$0.3) reinforcing the multiple images system hypothesis.

In addition to the detection arc, two out of the three other predicted images, z6.3 and radial image z6.4, were easily identified in both the HST and ALMA data. These images lie within $<0\farcs\,4$ of their predicted locations and have photometric redshifts similar to that estimated for MACS0600-arc. In the \textit{HST} data, image z6.3 consists of a bright "bulge" referred to here as z6.3a and an additional faint component z6.3b which, according to the three lens models, is likely the multiply-imaged counterpart of the arc, and on which the ALMA signal is centered. It is unclear whether image z6.3a is truly part of the lensed source despite not being quintuply imaged as well, and -- while we acknowledge the possibility that z6.3a may be related -- we consider only z6.3b a genuine multiple image of this system. In the ALMA data, z6.3b also shows a faint dust detection with a S/N$\sim$3.5 ($\rm{f_{peak}=226\pm63 \mu Jy.beam^{-1}}$ as measured with imfit). The 1.2mm flux ratio between 1.3b and the arc is 0.62$^{+0.33}_{-0.24}$ in excellent agreement with ratios observed in the near infrared and with the predicted lens-model magnifications (see below). [CII]158$\mu$m emission lines at $z=6.07$ are detected in all images of this system that are covered by ALMA's FOV: the arc, z6.3b and z6.4, strengthening further the adopted lensing configuration, again with line ratios that commensurate with the expectations from the lens models. More extensive details on these detections are given in a companion paper (Fujimoto et al., submitted).

While all models reassuringly agree on the positions of the five multiple images, there is larger scatter ($\sim$12$''$) in the predicted location of the fifth image, z6.5. The \textsc{GLAFIC} and \textsc{LTM} models predict image z6.5 lies outside the region covered by WFC3 data, making its identification more challenging, and two objects undetected on the ACS images but seen on the HAWK-I data are considered as potential candidates (z6.5a and z6.5b). The \textsc{Lenstool} model predicts the fifth image within the WFC3 field, and one clear dropout is identified as a potential candidate using the ACS, WFC3 and HAWK-I data (MACS0600-z6.5c). 

Magnification factors $\mu$ for the five images were estimated as follows. The two parametric models (\textsc{Lenstool} and \textsc{GLAFIC}) extracted values by first planting a compact source in the expected position, and adjusting its exact position with respect to the critical curves to match the observed [CII] 158 $\mu$m line flux ratio. The LTM magnification values were derived in a slightly different manner: here the source was not planted in the source plane but formed there by directly delensing the arc, and the magnification of the arc was then extracted by comparing the model's prediction for the three other multiple image sizes with their absolute magnification values. The magnification for the five images from the \textsc{GLAFIC} model is $\mu\sim33$ for the arc, $\mu\sim35$ for z6.3, $\mu\sim5.7$ for z6.4 and $\mu\sim14$ for z6.5. The \textsc{Lenstool} model yields $\mu\sim22.1$ for the arc, $\mu\sim14$ for z6.3, $\mu\sim2.1$ for z6.4 and $\mu\sim2.9$ for z6.5. The \textsc{LTM} model magnifications are $\mu\sim31.2$ for the arc, $\mu\sim13.5$ for z6.3, $\mu\sim2.1$ for z6.4 and $\mu\sim3.8$ for z6.5. We note that the magnifications for the arc are likely lower limits. If the magnification for the arc is instead estimated by directly averaging the magnification next to the critical curves, or alternatively by comparing the area of the arc to its delensed size, values around $\mu\sim150-200$ are obtained, although such estimates are typically more uncertain. 

We also take benefit from the detection of the dust continuum in two of the multiple images (z6.3 and the arc) to estimate the physical offset between the FIR continuum and the UV continuum. The position of the UV continuum was estimated on the F160W image using the centroid of the counterpart. In the image plan we measured an observed offset of 0$\farcs$24 and 0$\farcs$35 respectively for z6.3b and the arc. Accounting for the magnification, we estimate a physical offset of $<$0.4 kpc. 

\section{Spectroscopic follow-up }
\label{sec.spectro}
The detection of an emission line within the ALMA band 6 datacube suggests a redshift $z_{spec}$=6.07, assuming the line is [CII]158$\mu$m (Fujimoto et al. submitted). However, with only one line detected, a low/intermediate redshift interloper cannot be excluded. The relatively shallow MUSE data used to constrain the mass models did not show any features for the multiple-images of this system. Therefore we conducted a spectroscopic follow-up campaign with the Gemini Multi-Object Spectrograph (GMOS - \citealt{Hook2004}) installed on Gemini-North on three images of the system, namely MACS0600-arc, MACS0600-z6.3a and MACS0600-z6.4. Observations were done in service mode on the 18$^{th}$ and 19$^{th}$ October 2020 (ID : GN-2020B-Q-903 ; P.I. : A. Zitrin). We secured 4.5hrs  reaching a 1-sigma sensitivity of 8.9$\times$10$^{-19}$ erg/s/cm$^2$ over the wavelength range 505nm to 980nm. We reduced the data using the Gemini IRAF package, as recommended by the instrument team. We follow the standard reduction procedure including bias subtraction, flat fielding, wavelength calibration from the illumination of our mask by the CuAr lamp and flux calibration from the white dwarf Wolf 1346.

\begin{figure}
\includegraphics[width=9cm]{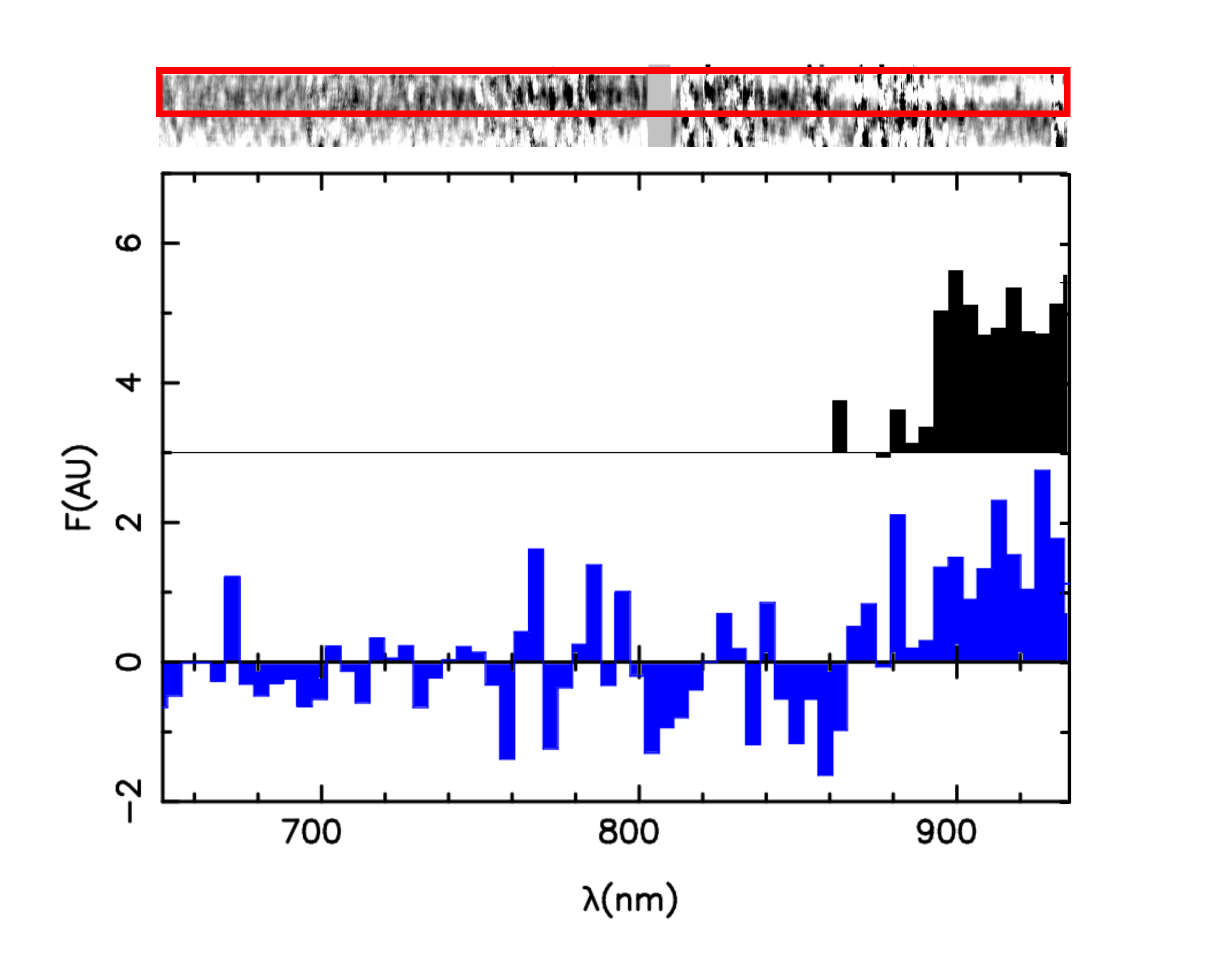}
\caption{\label{fig.GMOS} (\textbf{Top: }) 2D GMOS/Gemini spectrum of MACS0600-z6.3a smoothed with a boxcar of 5pixels. The red rectangle shows the position of our object in the slit. The continuum is clearly visible at the larger wavelength. (\textbf{Bottom: }) The blue line displays the extracted spectrum of MACS0600-z6.3a within a 1.5$\times$seeing diameter aperture. The grey spectrum is the best-fit of the \citet{Jones2012} spectra to our data. The best fit is found at $z_{spec}$=6.19$^{+0.06}_{-0.16}$.}
\end{figure}

Interestingly, the continuum of the more compact and brightest image (MACS0600-z6.3a) is visible on the Gemini spectra. We extracted the spectrum in a 1.5$\times$seeing diameter aperture and a break is clearly identified between 850nm and 900nm. To improve the signal, we binned the spectra in the spectral direction with a binning factor of 30 (see Figure~\ref{fig.GMOS}). We determined the spectroscopic redshift by fitting our observed spectra with a stacked spectra coming from 81 LBGs \citep{Jones2012}. We searched for the best-fit using a $\chi^2$ minimization technique over a redshift range between $z=$0 and $z=8$. The best fit is found at $z_{spec}$=6.19$^{+0.06}_{-0.16}$, consistent with both the photometric redshift and the [CII]158$\mu$m emission line detected in the ALMA data. One can also argue that if the break seen in the Gemini data is the Balmer Break at $z\sim$1.1 instead of the Lyman Break at $z\sim$6, we would have detected the [OII]3727,3729 doublet, which makes the high-redshift identification for this system robust. Moreover, after a careful visual inspection of the 2D spectra, no emission line is detected in any of the 3 images allowing us to place a firm upper limit on the Ly-$\alpha$ luminosity in this $z\sim$6 system at L(Ly-$\alpha$, 2$\sigma$) $<$ 2.4$\times$10$^{42}$ erg/s, corresponding to a rest-frame EW of $<$113\AA\ /$\mu$ and EW<3.6\AA\ assuming a magnification factor of $\mu$=31. 

\section{Physical Properties}
\label{sec.properties}

We determine the intrinsic (lensing-corrected) physical properties for those two images with a clear dust detection (MACS0600-arc and MACS0600-z6.3b) using \textsc{MAGPHYS} \citep{MAGPHYS}, which includes FIR models to fit the dust properties, while we use \textsc{BAGPIPES} \citep{BAGPIPES} to provide constraints for the remaining images. For the \textsc{BAGPIPES} runs,  we assume a constant Star Formation History (SFH) with a stellar mass ranging from 10$^6$ to 10$^{12}$ M$_{\odot}$, and an age ranging from 0.0 Gyr to the age of the Universe at $z=6.07$ . For each image we adopt the magnification from the \textsc{LTM} model, based upon the spectroscopic redshift, although as seen in the previous section the magnifications from the three models seem to agree fairly well. Reassuringly, we determine that all images have similar properties. Excluding  image MACS0600-z6.3a, which does not seem to contribute significantly to the flux compared to MACS0600-z6.3b, the mean stellar mass is 2.9$^{+11.5}_{-2.3}\times$10$^9$M$_{\odot}$, the mean SFR is 9.7$^{+23.0}_{-6.6}$M$_{\odot}$/yr and the mean dust mass (computed only from images MACS0600-arc and MACS0600-z6.3b) is 4.8$^{+4.5}_{-3.4}$ $\times$10$^6$M$_{\odot}$, where error bars take into account uncertainties on the magnification factor. The latter is obtained assuming a dust temperature of 30K, a dust emissivity of $\beta$=2.0 and the mean magnification. Including \textit{Herschel}/SPIRE 3$\sigma$upper limit(13.9mJy/$\mu$ at 250$\mu$m), we can rule out T$_{dust}\geq$85K (Sun et al. in prep).  We measure the UV slopes of all images from the best SED-fit and find an average $\beta$=$-$1.66$\pm$0.08. We follow the method described in \citet{Oesch2010} to determine the size of this galaxy, taking into account the \textit{SExtractor} half light radius and the \textit{Hubble} PSF. We apply this method on z6.4 since this image is not blended with other sources. After correcting for magnification, we find an intrinsic size of 0\farcs\,10 corresponding to  0.54$^{+0.26}_{-0.14}$ kpc at $z\sim$6.07, in good agreement with previous size estimates at these redshifts (\citealt{Bouwens2017}, \citealt{Kawamata2018}). The individual properties of each image are summarised in Table~\ref{properties}. 

\begin{table}
\hspace{-1.4cm} 
    \centering    

   \hspace{-1.4cm}  \begin{tabular}{l|ccccc} \hline
Source	&	M$_{\star}$	&	$\rm{M_{dust}}$	&	SFR		& $\beta$ &	$\mu$			\\
	    &	[$\times$10$^9$ M$_{\odot}$]	&	[$\times$10$^6$ M$_{\odot}$]			&	[M$_{\odot}$ / yr]			&	&				\\ \hline
	    
\small arc	&	2.2	$^{+	1.3	}_{-	0.9	} $ & 		4.5	$^{+	3.4	}_{-	3.1	} $  & 	6.1	$^{+	7.6}_{-	1.9	} $  & $-$1.6$\pm$0.1 & 	31	$^{+10}_{-9}$		\\
z6.3a	&	2.7	$^{+	1.9	}_{-	1.0	}$  & 		-		&	15.3	$^{+	4.6	}_{-	6.3	}$  & $-$1.6$\pm$0.2 & 	\textit{(13	$^{+3}_{-2}$)}		\\
z6.3b	&	1.1	$^{+	0.6	}_{-	0.5	}$  & 		5.1	$^{+	4.2	}_{-	3.2	} $  & 	5.9	$^{+	3.5}_{-	2.8	} $ &  $-$1.7$\pm$0.3 & 	13	$^{+3}_{-2}$		\\
z6.4	&	3.7	$^{+	2.3	}_{-	2.1	} $ & 		-	&	10.3$^{+	12.0	}_{-	6.0	}$ &  $-$1.7$\pm$0.1 & 	2.1	$^{+0.6}_{-0.7}$		\\
z6.5	&	4.7	$^{+	9.7	}_{-	3.4	} $ & 		-		& 10.7	$^{+	22.0	}_{-	5.8	}$ & $-$1.6$\pm$0.3 &	3.8	$^{+0.3}_{-0.4}$	\\ \hline
    \end{tabular}
    \caption{\label{properties} Physical properties of all multiple images of MACS0600-z6 computed with \textsc{MAGPHYS} \citep{MAGPHYS} for sources with dust detection (arc and z6.3b) and \textsc{BAGPIPES}  \citep{BAGPIPES} for the remaining. Uncertainties for the \textsc{LTM} model magnification represent the range of values predicted by four other trial LTM models. The central value is obtained by averaging the parameter value of the best fit-model obtained assuming the mean, min and max magnification value and the error bars included uncertainties on the magnification factor}
\end{table}
\begin{figure}
 \hspace{0cm}\includegraphics[width=10cm]{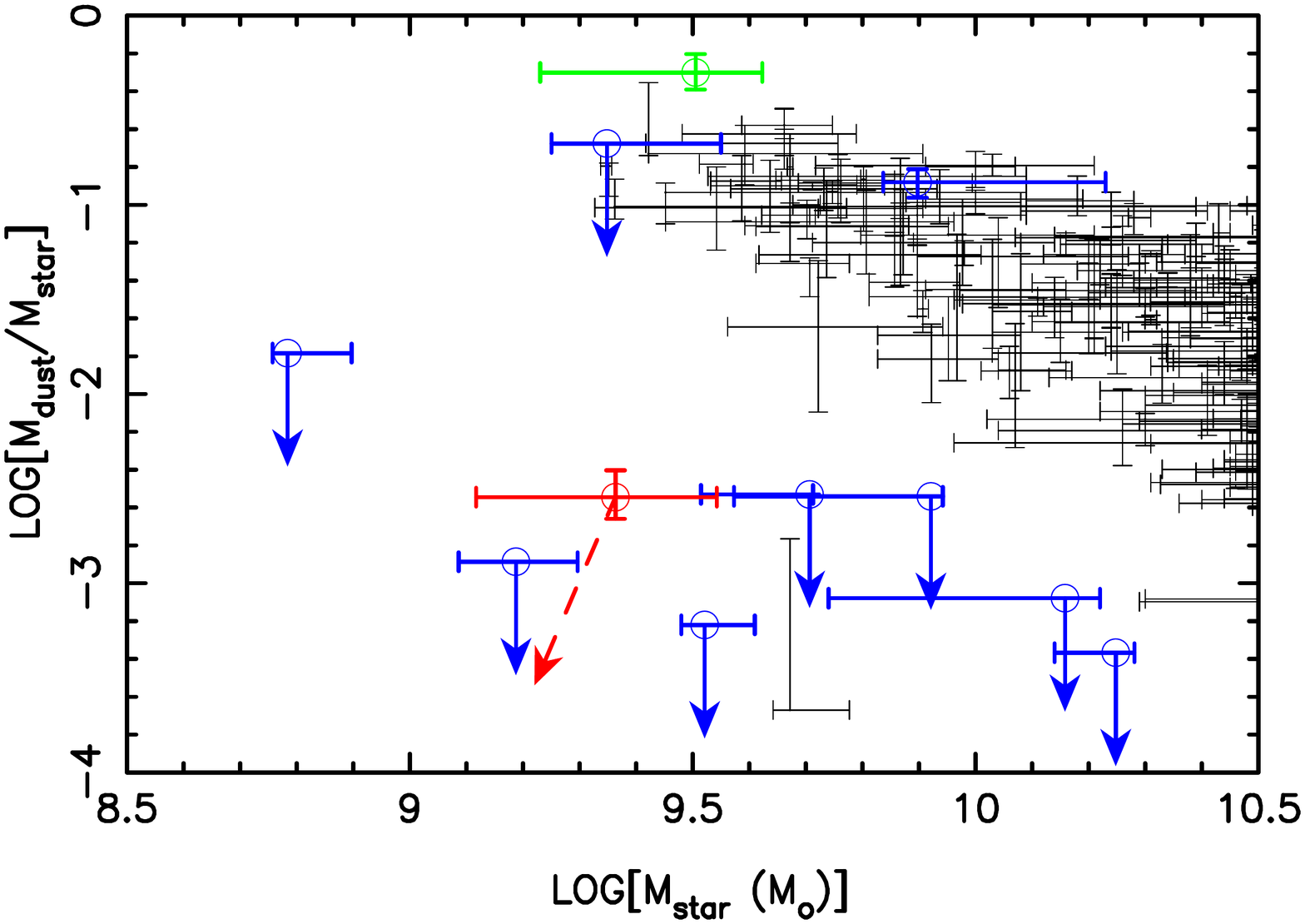}
\caption{\label{dust-to-mass} Dust to stellar mass ratio as a function of the stellar mass. The red dot shows the properties of MACS0600-z6, blue dots represent previously observed galaxies with dust constraints (detections or upper limits) with a spectroscopic redshift between $z\sim$6 and 7 (\citealt{Schaerer2015}, \citealt{Kanekar2013}, \citealt{Ouchi2013}, \citealt{Marrone2018}, \citealt{Bradac2017}, \citealt{Laporte2017_Frontier}, \citealt{Knudsen2016}, \citealt{Smit2018}). Black dots show lower redshift galaxies at $z\sim$1.5-3.5 (\citealt{Dudzeviciute2020},\citealt{Sklias2014}). The ratio for MACS0600-z6 is fairly high for its redshift, indicating either that it is relatively old or that its dust production was unusually efficient. The green circle shows the position of MAMBO-9, a sub-mm galaxy with an extreme dust to stellar mass ratio at $z=5.9$ \citep{Casey2019}; the distinct natures of these two galaxies (sub-mm vs LBG) makes a comparison of their dust production efficiency difficult. The red dashed arrow shows the direction in parameter space that the LBG would move when taking into account high dust temperature (T=85K) and uncertainties on the magnification..}
\end{figure}

We can compare the above results for MACS0600-z6 to similar spectroscopically confirmed galaxies at $z\sim$6-7 (either a detection or upper limit) in the literature (\citealt{Schaerer2015}, \citealt{Kanekar2013}, \citealt{Ouchi2013}, \citealt{Marrone2018}, \citealt{Bradac2017}, \citealt{Knudsen2016}, \citealt{Smit2018}). Most of these galaxies have upper limits on their dust content, emphasising the importance of MACS0600-z6.  Despite its modest stellar mass, its dust to stellar mass ratio is somewhat larger than that observed in other $z\sim$6 galaxies (Figure~\ref{dust-to-mass}). This implies either that MACS0600-z6 is already a relatively mature system or that its dust production was unusually efficient. We can provide an independent estimate of the stellar age via the amplitude of the Balmer Break (see e.g. \citealt{Scoville2016}). This requires comparing the fluxes in the F160W and 4.5$\mu$m bands, since the 3.6$\mu$m band could be contaminated by strong [O III] emission  \citep{Labbe2013}. Following the methodology discussed in \citet{Roberts-Borsani2020}, we estimate an age of 200$\pm$100 Myr, indicative of a formation redshift of $z\rm{_{form}}$=7.2$^{+0.8}_{-0.6}$, comparable to values estimated for the other $z\sim$6 galaxies cited above. This relatively young age would suggest that most of the dust detected by ALMA maybe produced by early core-collapse supernovae (SNe) \citep{Gall2011}. Assuming a Chabrier \citep{Chabrier2003} IMF for the production rate of SNe, for the measured dust mass and stellar age, each SN would have produced 0.3$^{+0.5}_{-0.2}$ M$_{\odot}$ of dust. Although this estimate is highly uncertain given the uncertain production rate of SNe in early metal poor galaxies and possible dust destruction and ejection processes in low mass galaxies at high redshift, it is perhaps somewhat larger than that derived locally ($<$0.2 M$_{\odot}$, see e.g. \citealt{Cherchneff2010}, \citealt{Indebetouw2014}, \citealt{Gomez2012}, \citealt{Gall2011}). Recent chemical evolution models predict dust-to-mass ratios similar to that observed for MACS0600-z6 could be achieved in $\leq$200 Myr \citep{Calura2017}. Even higher values have also been observed recently in MAMBO-9 \citep{Casey2019}, an intensely star-forming sub-mm galaxy whose properties are otherwise quite distinct from those of MACS0600-z6. Applying the same method, MAMBO-9 may be even more efficient to produce dust, but the different nature of these two objects makes the comparison of their dust production efficiency difficult.

\begin{figure}
\hspace{-0.5cm} \includegraphics[width=10cm]{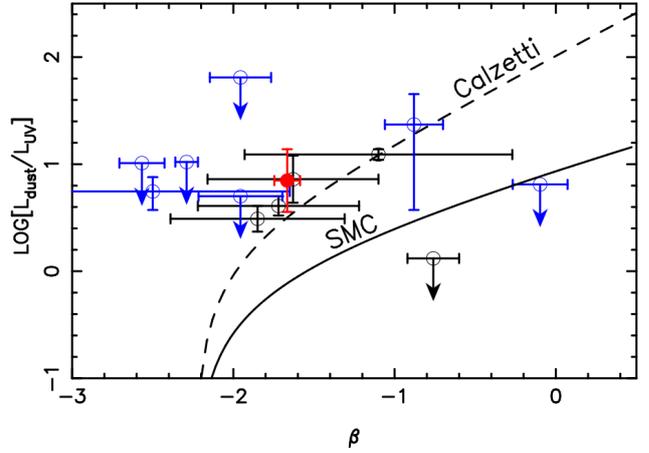}
\caption{\label{IRX-beta} The IR/UV luminosity ratio IRX as a function of the UV slope $\beta$. Black points represent galaxies at $z\geq$7, blue points are $z\sim$6 galaxies and the red circle shows MACS0600-z6. All values are computed assuming a dust temperature of 35K. The expected trend assuming the \citet{Calzetti2000} and SMC laws are shown.   }
\end{figure}

The ratio between the IR and UV luminosity, often referred to as IRX=$\log(L_{IR}/\rm{L_{UV}}$), can be compared to the UV continuum slope $\beta$ to offer insight into the dust extinction law at early cosmic epochs. \citet{Bouwens2016} found that $z>4$ galaxies with $\log[M{_{\star}}<$9.75] (comparable to MACS0600-z6) may have extinction laws that deviate from both the \citet{Calzetti2000} and SMC relations, in the sense of having lower IRX values. They argued that a higher dust temperature may explain this difference. Although there is some evidence for higher dust temperatures at $z\geq$8 (\citealt{Behrens2018}, \citealt{Laporte2019}, \citealt{Bakx2020}), there is currently no strong evidence for evolution at $z\sim$6 (e.g., \citealt{Schreiber2018};\citealt{Faisst2020}). Assuming a dust temperature of 35K to estimate the IR luminosity of MACS0600-z6 and the other $z\sim$6 galaxies, Figure~\ref{IRX-beta} shows that most $z\sim$6 galaxies follow the Calzetti law. 

\section{Discussion and Conclusion}

Recent studies have suggesteded that the dust temperature may be higher at high-redshift than what is measured at lower redshift. Obtaining an accurate measurement of the dust temperature in high-redshift galaxies requires to constrain the red slope and the peak of the FIR emission (see Figure 1 of \citealt{daCunha2008}).  The evolution of the dust temperature with redshift can be studied at intermediate redshift combining data from \textit{Herschel} and ALMA. \citet{Schreiber2018} demonstrate that the dust temperature goes from 25K at $z\leq$1 up to 40K at $z\sim$5.  At $z\geq$7 few sources have been observed at multiple sub-mm wavelengths (\citealt{Watson2015}, \citealt{Knudsen2017}, \citealt{Tamura2019}, \citealt{Bakx2020}, \citealt{Laporte2017}, \citealt{Laporte2019}). They all estimate a dust temperature at $z\geq$7 above the "standard" value of 30K observed at low-redshift, but deeper data at $\lambda \geq$1mm are needed. Recent theoretical models suggest that the dust temperature in $z\geq$5 galaxies is $T_{dust}$=45K for 0.1$\mu$m grain, and could reach as high temperature as $T_{dust}\geq$60K for smaller grains \citep{Ferrara2017}. In view of this, we explore whether a higher dust temperature would change our conclusions. For a temperature of 85K, the estimated dust mass would decrease to  $M_{dust}$=6.3$^{+1.6}_{3.6}$$\times$10$^6$M$_{\odot}$, and its dust-to-stellar mass ratio will be comparable to previous findings at high-redshift. Further ALMA observations at longer wavelength (e.g. in band 5) are needed to constrain the dust temperature and to refine our dust mass estimates. Moreover, adopting a higher dust temperature to study the IRX-$\beta_{uv}$ relation confirms our conclusion on the dust law preferred by our object, since an increase in dust temperature will increase the dust luminosity without changing the UV luminosity, and will therefore tend to higher IRX, consistent with the Calzetti law.

Such intrinsically faint sources ($M_{uv}$=-19.9$\pm$0.15) emitting light during the Epoch of Reionisation are considered to be dominant contributors to the cosmic reionisation of hydrogen (e.g. \citealt{Atek2015}, \citealt{Bouwens2017}). This strongly-lensed system has therefore the capabilities of testing this hypothesis. Recent studies of the Ly-$\alpha$ emission line  (either from LBG or LAE) using MUSE/VLT  (e.g. \citealt{deLaVieuville2019}, \citealt{deLaVieuville2020}) or the Hyper Suprime-Cam/Subaru (e.g. \citealt{Konno2018}) show that the luminosity of Ly-$\alpha$ at $z\ge$5.5 ranges from 10$^{40}$ to 10$^{43.8}$ erg/s. The Ly-$\alpha$ properties in MACS0600-z6 (L(2$\sigma$)$<$2.4$\times$10$^{42}$ erg/s and EW$_{RF}<$113 \AA\ ) are therefore not particularly exceptional and comparable in luminosity and Ly-$\alpha$ EW to the bulk of the galaxy population at $z\sim$6.

We report the discovery of a $z=$6.07 dusty galaxy strongly lensed by the cluster MACS0600.1-2007 comprising 5 multiple images. We used three different mass models for the cluster, showing good agreement between predicted magnification factors and observed flux ratio between the five images of this galaxy.  The physical properties we deduced from a detailed SED analysis show that this galaxy has a stellar mass of $\sim$ 10$^9$ $\rm{M_{\odot}}$, a dust mass of  $\sim$10$^6$ $\rm{M_{\odot}}$ respectively, a small star formation rate ($<$10$\rm{M_{\odot}.yr^{-1}}$) and a size ($\sim$0.5 kpc) consistent with previous observations. We estimate the dust production rate in this galaxy and conclude that it seems higher than what has been observed in $\rm{z\leq5}$ galaxies. Further sub-mm observations, as provided by ALMA, are crucial to improve the estimate of the dust content of this galaxy.

\section*{Acknowledgements}
We thank the referee for providing useful comments which improved the quality of this paper, the Gemini Helpdesk team for their help with the reduction of Gemini data and Ian Smail for useful comments on this manuscript.
NL acknowledges support from the Kavli Foundation. RSE acknowledges funding from the European Research Council (ERC) under the European Unions Horizon 2020 research and innovation program(grant agreement No. 669253). MO is supported by World Premier International Research Center Initiative (WPI Initiative), MEXT, Japan, as well as KAKENHI Grant-in-Aid for Scientific Research (A) (17H01114, 19H00697, and 20H00180) through Japan Society for the Promotion of Science (JSPS). KK and TW are supported by JSPS KAKENHI Grant Number JP17H06130 and by the NAOJ ALMA Scientific Research Grant Number 2017-06B. FEB acknowledges support from ANID-Chile Basal AFB-170002, FONDECYT Regular 1200495 and 1190818, and Millennium Science Initiative ICN12\_009. KKK acknowledges support from the Knut and Alice Wallenberg Foundation. Y.A. acknowledges support by NSFC grant 11933011. GB and KC acknowledge funding from the European Research Council through the award of the Consolidator Grant ID 681627-BUILDUP. 

This paper makes use of the following ALMA data: ADS/JAO.ALMA\#2018.1.00035.L. ALMA is a partnership of ESO (representing its member states), NSF (USA) and NINS (Japan), together with NRC (Canada), MOST and ASIAA (Taiwan), and KASI (Republic of Korea), in cooperation with the Republic of Chile. The Joint ALMA Observatory is operated by ESO, AUI/NRAO and NAOJ. This work is based on observations taken by the RELICS Treasury Program (GO 14096) with the NASA/ESA HST, which is operated by the Association of Universities for Research in Astronomy, Inc., under NASA contract NAS5-26555. Based on observations obtained at the international Gemini Observatory, a program of NSF's OIR Lab, which is managed by the Association of Universities for Research in Astronomy (AURA) under a cooperative agreement with the National Science Foundation on behalf of the Gemini Observatory partnership: the National Science Foundation (United States), National Research Council (Canada), Agencia Nacional de Investigaci\'on y Desarrollo (Chile), Ministerio de Ciencia, Tecnolog\'ia e Innovaci\'on (Argentina), Minist\'erio da Ci\^encia, Tecnologia, Inovac\~oes e Comunicac\~oes (Brazil), and Korea Astronomy and Space Science Institute (Republic of Korea).

\section*{Data Availability}
The data underlying this article will be shared on reasonable request to the corresponding author.




\bibliographystyle{mnras}
\bibliography{arxiv} 




\begin{landscape}

\begin{table}

   \hspace{-0.5cm}  \begin{tabular}{l|cc|cccccccccc|c} \hline
    ID & RA & DEC & F435W & F606W  & F814W & F105W & F125W & F140W & F160W & K$_s$ & 3.6$\mu$m & 4.5$\mu$m & $\rm{z_{phot}}$\\ \hline
    Foreground   & 06:00:09.13 & $-$20:08:26.67 &$>$27.34  & 26.55$\pm$0.35 & 24.89$\pm$0.11 & 25.01$\pm$0.02 & 25.27$\pm$0.15 & 24.71 $\pm$0.03 & 24.67 $\pm$0.01 & 23.64$\pm$0.14 & 23.90$\pm$0.27  & 24.70$\pm$0.62 & 4.4$^{+0.4}_{-0.4}$ \\
    MACS0600-arc $^*$ & 06:00:09.13 & $-$20:08:26.59 & $>$27.34  & $>$26.74 & $>$26.35 & 23.83$\pm$0.14 & 24.08$\pm$0.16 & 23.89 $\pm$0.15 & 23.39 $\pm$0.15 & $>$23.5 & 22.30$\pm$0.08  & 22.75$\pm$0.13 & 6.7$^{+0.2}_{-0.7}$ \\ \hline
    MACS0600-z6.3a $^*$ & 06:00:09.56 & $-$20:08:10.80 & $>$27.9 & $>$29.2 & 26.46$\pm$0.31 & 24.12$\pm$0.12 & 24.10$\pm$ 0.13 & 24.10$\pm$0.13 & 23.86 $\pm$0.11 & $>$23.5 & 22.66$\pm$0.11 &	23.16$\pm$0.12 & 6.7$^{+0.2}_{-0.6}$ \\
    MACS0600-z6.3b  & 06:00:09.53 & $-$20:08:10.84 & $>$26.8 & $>$27.8 & $>$ 27.5 & 25.24$\pm$0.13 & 24.92$\pm$0.13 & 25.01$\pm$0.14 & 24.88$\pm$0.12 & $>$23.5 & 23.66$\pm$0.29 &	24.16$\pm$0.29  & 6.8$^{+0.5}_{-1.2}$\\
    MACS0600-z6.4 $^*$ & 06:00:08.57 & $-$20:08:12.30 & $>$28.1	& $>$ 29.4 & 27.11$\pm$0.25 &	25.51$\pm$0.26 & 26.25$\pm$0.31 & 	25.96$\pm$0.24 & 25.75$\pm$0.23 & $>$24.6 & $>$24.2 & $>$ 23.6 & 5.7$^{+0.6}_{-0.6}$     \\
    MACS0600-z6.5a$^{ \dagger }$ &  06:00:05.16 & $-$20:07:13.87 & $>$27.3 & $>$28.4 & $>$28.0 & - & - & - & - & 25.47$\pm$0.31 & $>$24.8 & $>$ 24.0 & - \\ 
    MACS0600-z6.5b$^{ \dagger }$ & 06:00:04.49 & $-$20:07:14.93 & $>$27.5 & $>$27.8 & $>$28.1 & - & - & - & - & 25.17$\pm$0.24 & $>$25.5 & $>$ 24.2 & -\\
    MACS0600-z6.5c & 06:00:05.57 & $-$20:07:20.84 & $>$27.5 & $>$27.8 & 25.86$\pm$0.05 & 24.87$\pm$0.02 & 24.73$\pm$0.02 & 24.66$\pm$0.02 & 24.65$\pm$0.02 & 24.37$\pm$0.20 & $>$25.5 & $>$ 24.2  & 5.7$^{+0.2}_{-0.2}$ \\
    \hline

    \end{tabular}
    \caption{\label{photometry} Photometry of the multiple images of this $z\geq$6 system. Upper photometric limits on the ACS and WFC3 data were estimated in a 0.4$''$ radius aperture, on HAWK-I data in a 0.8$''$ radius aperture (taking into account the seeing of the data), and in a 1.2$''$ radius aperture for IRAC images. For the photometric extraction of z6.3a and z6.3b, we used GALFIT on ACS and WFC3 images assuming a Sersic profile for both. On the IRAC data, where the separation is not clearly resolved, we used the flux ratio observed at 1.6$\mu$m to separate the contribution of each source. 
    \newline
    $^{\star}$ spectroscopically confirmed at $z=6.07$ (see Fujimoto et al. submitted)
    \newline
    $^{\dagger}$ Candidates for the fifth image of this system. The absence of WFC3 data at the expected position makes difficult the clear identification of the image.}
\end{table}
\bsp	
\label{lastpage}
\end{landscape}


\end{document}